\documentclass[aip,reprint]{revtex4-1}

\draft 

\usepackage{graphicx}
\usepackage{dcolumn}
\usepackage{bm}
\usepackage{booktabs}
\usepackage{lineno}
\usepackage{amsmath}
\usepackage{tabu}
\begin{document}


\title{ Simultaneous $ \chi^{(2)} $- $ \chi^{(2)} $ and  $ \chi^{(2)} $-$ \chi^{(3)} $ nonlinear processes generation in thin film lithium tantalate microcavity } 



\author{Xiongshuo Yan*}
\affiliation{State Key Laboratory of Advanced Optical Communication Systems and Networks, School of Physics and Astronomy, Shanghai Jiao Tong University, Shanghai 200240, China}
\author{Miao Xue*}
\affiliation{State Key Laboratory of Advanced Optical Communication Systems and Networks, School of Physics and Astronomy, Shanghai Jiao Tong University, Shanghai 200240, China}

\author{Jiangwei Wu}
\affiliation{State Key Laboratory of Advanced Optical Communication Systems and Networks, School of Physics and Astronomy, Shanghai Jiao Tong University, Shanghai 200240, China}

\author{Rui Ge}
\affiliation{State Key Laboratory of Advanced Optical Communication Systems and Networks, School of Physics and Astronomy, Shanghai Jiao Tong University, Shanghai 200240, China}

\author{Tingge Yuan}
\affiliation{State Key Laboratory of Advanced Optical Communication Systems and Networks, School of Physics and Astronomy, Shanghai Jiao Tong University, Shanghai 200240, China}

\author{Yuping Chen}
\email{ypchen@sjtu.edu.cn}
\affiliation{State Key Laboratory of Advanced Optical Communication Systems and Networks, School of Physics and Astronomy, Shanghai Jiao Tong University, Shanghai 200240, China}

\author{Xianfeng Chen}
\email{yxfchen@sjtu.edu.cn}
\affiliation{State Key Laboratory of Advanced Optical Communication Systems and Networks, School of Physics and Astronomy, Shanghai Jiao Tong University, Shanghai 200240, China}
\affiliation{Shandong Normal University,Collaborative Innovation Center of Light Manipulation and Applications, Jinan, 250358, China}



\date{\today}

\begin{abstract}
On-chip efficient nonlinear functions are instrumental in escalating the utilities and performance of photonic integrated circuits (PICs), especially for a wide range of classical and quantum applications, such as tunable coherent radiation, optical frequency conversion, spectroscopy, quantum science, etc. Lithium tantalate (LT) has been widely used in nonlinear wavelength converters, surface acoustic wave resonators, and electro-optic, acoustic-optic devices owing to its excellent optical properties. 
Here, we fabricated a Z-cut lithium tantalate on insulator (LTOI) microdisk with high quality(Q) factors in both telecom (10$^{6}$) and visible (10$^{5}$) bands by optimizing the fabrication. 
With the Q factor of the LTOI microdisk increasing, we can obtain higher pump light intensity in the cavity which is beneficial to get more optical nonlinear effect easily.
By making use of the mode phase matching of interacting waves and inputting high pump power, we experimentally observed on-chip near-infrared light, visible (red, green), and ultraviolet (UV) from
microresonator-based $ \chi^{(2)}-\chi^{(2)}$, $ \chi^{(2)}-\chi^{(3)}$, and $\chi^{(2)}$ nonlinear processes such as cascaded four-wave mixing (cFWM), cascaded sum-frequency generation
(cSFG), third harmonic generation (THG), second harmonic generation (SHG).
It is believed that the LTOI can support a variety of on-chip optical nonlinear processes, which heralds its new application potential in integrated nonlinear photonics.

\end{abstract}


\maketitle 

On-chip efficient and compact frequency converters are important and promising to be one of the key components in the wavelength division multiplexing (WDM) systems and have a range of applications, such as tunable coherent radiation [1,2], telecommunications [3,4], spectroscopy, and quantum optics [5,6,7,8].
Nonlinear optics provides a convenient route to access the frequencies we are interested in.  Since the nonlinear susceptibilities are very small as $ \chi^{(2)}\sim10^{-12} $m/V, $ \chi^{(3)} \sim 10^{-24}$ m$^{2}$/V$^{2}$ [9], the nonlinear response of a material system can be obtained only when laser light is sufficiently intense and the material should better have large nonlinear susceptibilities. That means to achieve efficient nonlinear reactions, we should load the laser with enough intensity and choose these materials with large nonlinear susceptibilities.
Optical whispering 
gallery mode (WGM) microcavities owing to their strong confinement of a light field (ultrahigh quality Q factors and small mode volumes) could considerably enhance the light-matter interaction [10], making it an ideal platform for studying a broad range of nonlinear optical effects [11-14], ultrahigh sensitivity sensor [15,16], and other physical applications [17,18]. It means that by making use of an optical microcavities structure, we could obtain intense laser under a small input. 
It should be noted that although the structure of optical WGM microcavities can enlarge the intracavity laser intensity, it also requires two criteria as phase matching and a large mode overlap among the interaction modes to obtain effective nonlinear processes [19,20].

Lithium niobate (LN), as one of the famous nonlinear materials, has large second-order nonlinear susceptibility and many other excellent physical properties such as high refractive index, strong electro-optic, acousto-optic, and piezoelectric effects, and has become one of the research hotspots recently [21]. 
Many on-chip photonic devices such as quantum transducers, modulators, filters, etc [22-25], and nonlinear processes (e.g. SHG, OPO, frequency combs, etc [26-40]), based on LN have been developed. What's more, rare-earth ion-doped thin film LN is also achieved, and on-chip microcavity lasers and waveguide amplifiers are investigated, although LN itself has no gain characteristics [41-46].
Nevertheless, LN has the advantage of a large nonlinear coefficient for nonlinear processes, there are some shortages for LN such as low optical damage threshold, and UV band opacity (0.4-5.5 $\mu$m) [47], which limits its application in high power light-input conditions, especially for multi-nonlinear processes in ultra-high-Q microcavities and the mid-UV [48,49].

\begin{figure*}[]
	\centering
	\includegraphics[width=\linewidth]{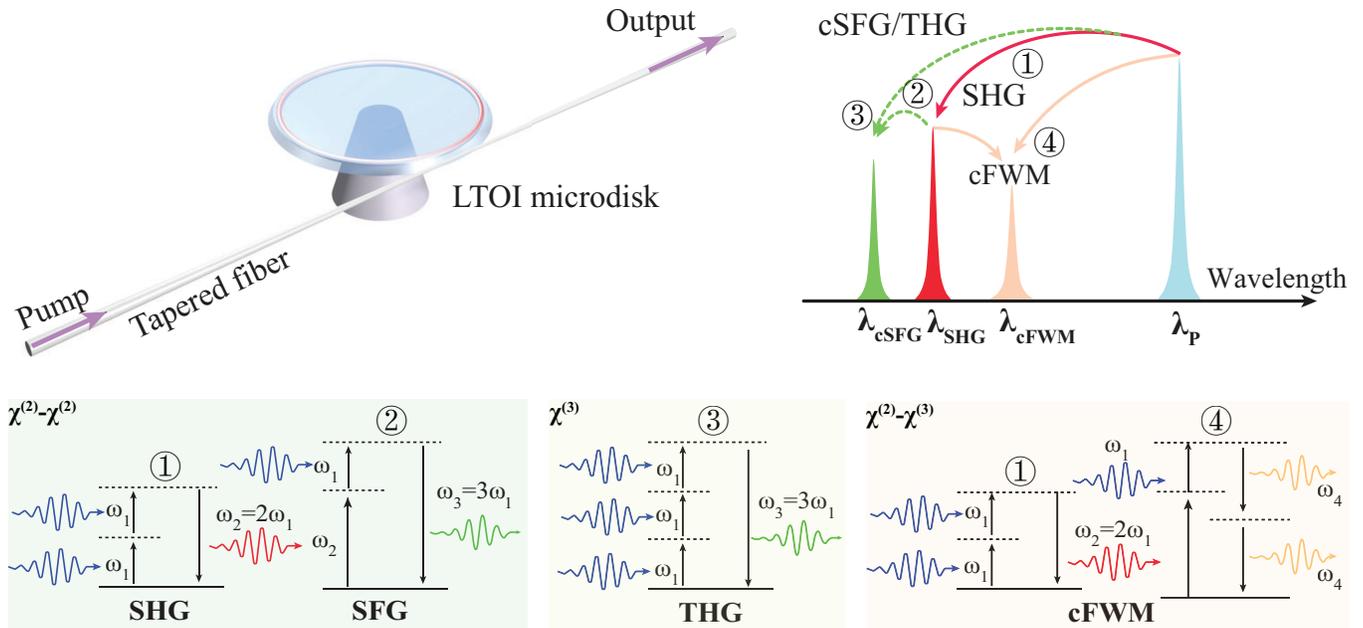}
	\caption{ Schematic of the multiple nonlinear processes generated in the LTOI microdisk  and corresponding energy level diagrams.}
	\label{Figure1}
\end{figure*}
Lithium tantalate (LT) as one of the important nonlinear nanophotonic platforms has analogous physical properties to LN [50]. Compared to the LN, LT has a higher laser-radiation-induced damage threshold, higher photorefractive damage threshold, and a wider UV transparency (0.28-5.5 $\mu$m). What's more, the LT can withstand large input power, which is beneficial to get more efficient multiple nonlinear optical processes.
There is also much research work about the LT based on its similar physical properties such as
proton exchange and femtosecond laser direct writing waveguides [51-53], periodically poled lithium tantalate (PPLT), and nonlinear frequency generation [54-68], entangled photons [69], holographic storage [70,71], strong coupling resonator [72], microwave communication [73], SAW resonators, and modulators [74,75], etc.
However, most of this research is based on LT bulk materials, then the structure is very large and difficult to integrate.  
Considering the excellent optical nonlinearity of LTOI, it may pave the way to achieve an on-chip wide range of wavelengths coverage via direct optical transition which is generally considered challenging [76].

In this work, we fabricated an LTOI microdisk resonator with a high Q factor and experimentally observed multiple nonlinear process generations in it. On-chip near-infrared, visible (red, green), and UV light are obtained, which come from the cascaded four-wave mixing (cFWM, third-order nonlinearity), second harmonic generation(SHG), cascaded sum-frequency generation (cSFG, second-order nonlinearity),  respectively. The nonlinear optical processes generated in the LTOI microdisk are shown in Fig. 1. Our work shows a way in the use of on-chip nonlinear photonics based on LTOI to access interested wavelengths in the near-infrared visible and UV spectrum. 
	
The device, 50-$\mu$m-diameter microdisk resonators, is fabricated on a
Z-cut LTOI wafer which has a sandwich structure, from top to bottom as 600-nm-thickness LT crystal, 2-$\mu$m-thickness silica, and 500-$\mu$m-thickness silicon substrate (shown in Fig. 2(a), prepared by NanoLN inc.). The main fabrication processes are as follows: First, the disk periphery
is milled precisely at the corner edges of LTOI (for better coupling) by using the focused ion beam
(FIB) and scanning electron microscope (SEM) dual beam system (ZEISS Auriga). To achieve a
smoother sidewall of the microdisk resonators, we optimized the current and dose of the ion beam.  A microring scanning pattern with a smaller current and dose of ion beam  was also used to remove residual
the LT crystal from the edge of the disk with smaller beams of gallium ions, which can reduce the scattering loss. 
\begin{figure}[]
	\centering
	\includegraphics[width=\linewidth]{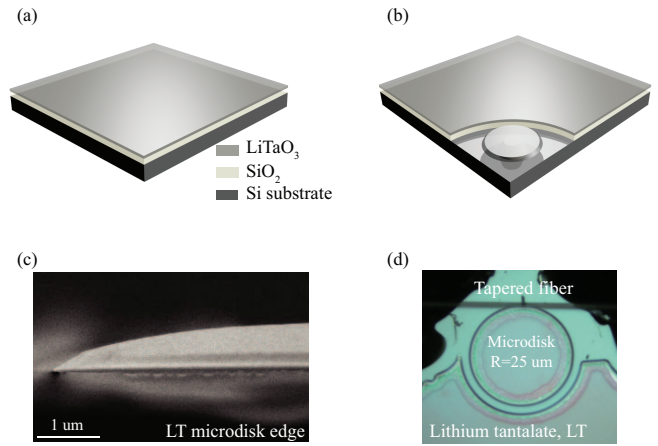}
	\caption{(a) LTOI sample,  (b) The schematic
		LTOI microdisk with silica pedestal, (c) and  (d) are the scanning electron microscope (SEM) image of the LTOI microdisk edge and optical microscope image (top view) of the whole LTOI microdisk, respectively.}
	\label{Figure1}
\end{figure}
What's more, the sample was slightly polished by chemo-mechanical polishing to further smooth the disk periphery.
At last, the silica layer underneath the LT was partially etched to form the
supporting pedestals and support the microdisk by immersing it in buffered oxide etching (BOE)
solution.
Fig. 2(b) is the final shape of the LTOI microdisk. The outermost ring with a perfectly round shape is the boundary of the
LTOI microdisk. The inner nearly-round border is the boundary of the silica pillar, which roundness is caused by the isotropic corrosion of silica by BOE. 
Fig. 2(c) and (d) are the scanning electron image (SEM) of the LTOI microdisk resonator edge and optical microscope image of the whole LTOI microdisk with a tapered fiber from the top view, respectively.
\begin{figure}[]
	\centering
	\includegraphics[width=\linewidth]{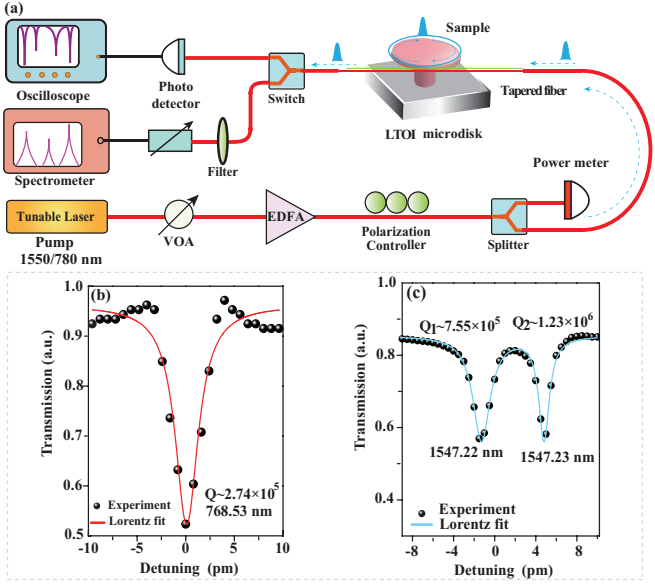}
	\caption{ (a) Experimental setup, VOA: variable
		optical attenuator, EDFA: erbium-doped fiber amplifier. (b), (c) are the Lorentzian fitting of a
		measured mode around 768.53 nm and 1547.23 nm, respectively.}
	\label{Figure1}
\end{figure}

The experimental setup is shown in Fig. 3(a).
Narrow-linewidth tunable continuous wave laser (New Focus TLB-6728 (telecom), TLB-6712 (visible), linewidth$ < $200 kHz) are used as the pump light sources. The pump laser power can be adjusted by a variable optical attenuator (VOA) at a suitable value, then it can be amplified by an erbium-doped fiber amplifier (EDFA). After that, the polarization of the input light is controlled by a polarization controller (PC), then
followed by a 99:1 single-mode fiber splitter for the wave splitting. To monitor the input
power in real time, the 1\% port is connected to a power meter. The pump power in the 99\% port is
launched into a tapered fiber to couple light into the LTOI microdisk. The tapered fiber with a waist
diameter of approximately 1$ \sim $2 µm is made by the heating and pulling method and placed on a precise
three-dimensional (3D) piezo nanostage. That means, we can flexibly change the polarization of
the input laser and the gap between the tapered fiber and the LTOI microdisk, which is related
to coupling loss, to optimize the coupling efficiency and make an efficient nonlinear process a
reality. The input laser and generated emission from the LTOI microdisk, monitored by an optical
microscope (not shown in Fig. 3(a)) from the top view, are also coupled out by the same tapered
fiber, then followed by a switch. One path is connected to an InGaAs photodetector (PD) and
an oscilloscope (OSC)to monitor the transmission spectrum. Another path is linked to an FC/PC
collimator followed by a short pass filter (Thorlabs, Inc., Model: FES0450, 450 nm cutoff) and
the optical spectrum analyzer (OSA, Ocean Optics, Inc., Model HR2000, detection range: 200 to
1100 nm) to monitor the generated spectrum.

To characterize the Q factor of the LTOI microdisk, we input small optical power both in visible and telecom bands into the cavity to avoid cavity thermal effects but also keep it enough to measure the transmission spectrum of the device. 
The whispering-gallery-mode
(WGM) in the visible band with the central wavelength at 768.53 nm shows the loaded optical Q-factor is  $2.74\times10^{5}$, as shown in Fig. 3(b). In the telecom band around 1547.22 nm, the loaded optical Q-factor of two WGM modes is $7.55\times10^{5}$ and $1.23\times10^{6}$ shown in Fig. 3(c), respectively. The device with a high Q factor 
exhibits very low optical losses in the LTOI microdisk. 

\begin{figure}[]
	\centering
	\includegraphics[width=\linewidth]{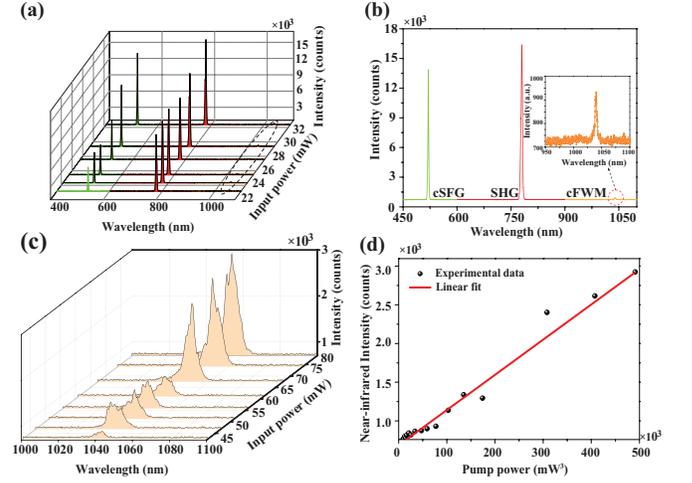}
	\caption{(a) The recorded spectrum of the visible signal (SHG, cSFG/THG). (b) The recorded spectrum of the SHG,
		cSFG/THG, and cFWM signal with the input pump light in the telecom band (36.19 mW). (c) cFWM near-infrared signal generation with different input power. (d) near-infrared signal intensity with different pump power.  }
	\label{Figure1}
\end{figure}

When the LTOI microdisk is pumped in the telecom band around 1556.86 nm, the visible signals (red, 778.43 nm, and green 519.03 nm) generation are observed at first, which originates from the SHG and cSFG/THG processes, respectively. Fig. 4(a) shows the visible signals observed by the optical spectrum analyzer at different input powers in the device. With the increasing pump power, we found that a signal in the near-infrared band (around 1037.75 nm) was generated (shown in Fig. 4(a) black dotted ellipse). We focused on the near-infrared signal firstly.
Fig. 4(b) shows the recorded spectrum under the input pump power at 36.19 mW, the near-infrared signal becomes apparent (red dotted circle). Fig. 4(c) shows the power dependence of the near-infrared signal on the different fundamental pumps in the telecom band, by keeping increasing the input power to around 78.89 mW. Fig. 4(d) shows the near-infrared signal intensity upon the different pump power. The cubic relationship between the near-infrared signal and the pump power is also in good agreement with the experimental data. The near-infrared signal was observed when the input pump power was greater than around 23 mW. 
For these nonlinear processes, the Hamiltonian of the multiple nonlinear processes can be written as [77],
\begin{eqnarray}
	\begin{aligned}
		\textbf{H} &= (\omega_{a,0}-\omega_{a})a^{\dag}a+(\omega_{b,0}-\omega_{b})b^{\dag}b+(\omega_{c,0}-\omega_{c})c^{\dag}c\\
		&+(\omega_{d,0}-\omega_{d})d^{\dag}d+g_{21}(a^{\dag2}b+a^{2}b^{\dag})+g_{22}(abc^{\dag}+ca^{\dag}b^{\dag}) \notag  \\
		&+g_{31}(a^{\dag3}c+a^{3}c^{\dag})+g_{32}(a^{\dag}b^{\dag}d^{2}+abd^{\dag2})+\varepsilon_{a}(a+a^{\dag})
	\end{aligned}
\end{eqnarray}
where a, b, c, d are the resonant mode of the fundamental mode, the second-harmonic mode, the third-harmonic mode, and the cascade four-wave mixing mode, respectively. $\omega_{i,0}-\omega_{a}$, (i= a, b, c) represents the detuning of the resonant mode to the photon frequency. $g_{21}$, $g_{22}$, $g_{31}$, $g_{32}$ represent the single-photon coupling
strength of SHG, SFG, THG, and cFWM generation. $\varepsilon_{a}$ represents the pump field strength. The coupled mode equations of these interaction modes can be obtained by following the Heisenberg equation and mean-field approximation.

It is worth mentioning that the
cFWM in the visible band is usually neglected for the infrared band pump light and the low Q of visible modes in the microcavity [77]. Combining our experimental parameters, we firstly calculated SHG intensities for different Q values (10$^{5}$ and 10$^{6}$) under the assumption that the phase matching condition is satisfied, shown as Fig. 5(a). We found that the SHG intensity of the LTOI microdisk Q factor with 10$^{6}$ is one order of magnitude ($ \sim $16) larger than the microdisk with the Q around 10$^{5}$. The large SHG intensity may be more conducive to the generation of near-infrared signals. A high Q-factor of microdisk enhances light-matter interactions, making it easier to generate nonlinear processes in LTOI microdisk resonators. 
\begin{figure}[htbp]
	\centering
	\includegraphics[width=\linewidth]{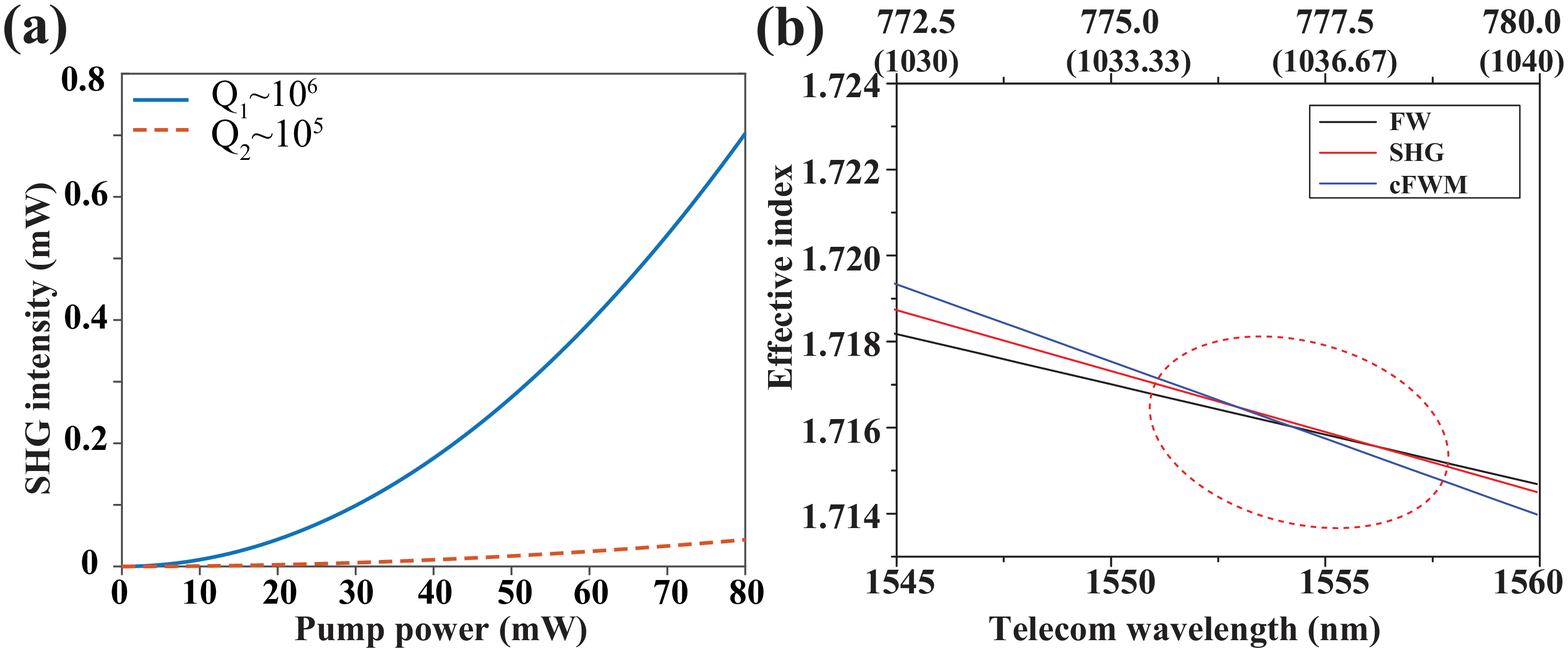}
	\caption{(a) visible SHG intensity with different pump power and Q factors in the LTOI microdisk.
		(b) Effective indices as functions of wavelength in the telecom (FW), visible (SHG) and near-infrared (cFWM).}
	\label{Figure5}
\end{figure}
Then, we analyzed the origination of the near-infrared signal and found that $\omega_{p}+\omega_{SHG}=2\omega_{NIF}$, where $\omega_{p}$, $\omega_{SHG}$, and $\omega_{NIF}$ are the frequency of the pump, SH signal, and near-infrared signal, respectively. And we further calculated the effective refractive indices of the input pump wave (FW), its SH signal (SHG), and the generated near-infrared signal (cFWM) as shown in Fig. 5(b), respectively. From Fig. 5(b) we observed that the mode phase matching can be satisfied too.
That means the near-infrared signal originated from a cFWM nonlinear process in the LTOI microdisk resonator, which is a $ \chi^{(2)} $-$ \chi^{(3)} $ nonlinear optical process. 
It should be noted that as we continue to increase the pumps power in the telecom band, the wavelength of the generated near-infrared signal fluctuated slightly, which should be caused by the thermal effect and photorefractive effect under the high input power condition.

\begin{figure}[htbp]
	\centering
	\includegraphics[width=\linewidth]{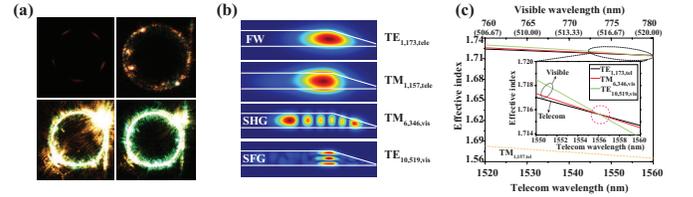}
	\caption{(a) Optical image of the generated visible signal.
		(b) Simulated mode profiles.  (c) Effective indices as functions of wavelength of TE$_{1,173,tele}$, TM$_{1,157,tele}$ in the telecom and TM$_{6,346,vis}$, TE$_{10,519,vis}$. }
	\label{Figure5}
\end{figure}

For the visible signal, it even became visible to the naked eye under the high input pump power.
Fig. 6(a) shows the optical image of the scattered visible signal generated in the LTOI microdisk.
We know that phase matching in the LTOI microdisk is needed to obtain effective nonlinear processes. 
Then we analyzed the phase-matching condition in the device.
Considering our microdisk fabricated on a 600-nm Z-cut LT without periodic polarization, 
we simulated possible mode profiles of the pump and visible signal in the device shown in Fig. 6(b) where we use indices $n$ and $m$ to denote radial and azimuthal mode numbers for the TE/TM$_{n,m}$ profiles in the microdisk. Combining the mode profiles and their effective refractive indices, the phase matching condition is shown as mode phase matching (MPM) at around 1556 nm, shown in Fig. 6(c) where the inset is a partial enlargement of the TE$_{1,173,tele}$ (FW in the telecom band), TM$_{6,346,vis}$ (SHG signal, red, around 778.63 nm), TE$_{10,519,vis}$ (cSFG signal, green, around 519.03 nm) effective refractive indices, respectively. 
From the Fig. 6(c) we can find that the effective refractive indices of mode profiles (TE$_{1,173,tele}$, TM$_{6,346,vis}$, TE$_{10,519,vis}$) are very close in the 1556-1560 band, and the closest phase matching area around 1556 nm at the temperature of 25 °C. That is in good agreement with our experimental results. 

It is worthy of note that generally speaking, the efficiency of the
third-order nonlinear process is very low, especially for the very weak $ \chi^{(3)} $ susceptibilities. Of course, one can design the dispersion, mode overlap area, and domain structure carefully to obtain high conversion efficiency. However, for these multiple nonlinear processes, more factors need to be considered and more research is needed. For example, for the nonlinear process of initial visible light generation, there consist of three nonlinear processes SHG, cSFG, and THG. The THG signal conversion is dominated by the cascaded SHG-SFG when the pump power is low and dominated by the direct THG under high pump conditions [78]. What's more, for the nonlinear processes in a microcavity, the cavity resonance and the external coupling conditions also have an important effect on conversion efficiency.

Although we have achieved the nonlinear processes in the LTOI microdisk, we investigated the on-chip UV generation based on the advantage of LT's UV transparency by loading the visible pump. The on-chip UV generation is based on the SHG nonlinear process. Fig. 7(a) shows the UV signal (384.3 nm) at different visible input power. The UV generation processes in the LTOI microdisk can be described with the coupled-mode equation,
\begin{eqnarray}
	\begin{aligned}
		\frac{da}{dt}&=-\frac{\alpha_{a}}{2}a-i\bigtriangleup_{a}a + i\omega_{a}\kappa L a^{*}b + \sqrt{\eta} a_{in},\\
		\frac{db}{dt}&=-\frac{\alpha_{b}}{2}b-i\bigtriangleup_{b}b + \frac{i}{2}\omega_{b}\kappa L a^{2}, 
	\end{aligned}
\end{eqnarray}
where a, b are the intracavity field amplitudes of $\omega_{a}$ and $\omega_{b}  $, ($ \omega_{b} = 2 \omega_{a} $) respectively. $ \alpha_{a} $, $ \alpha_{b} $ are the respective loss rates. $ \bigtriangleup_{a} =  \varpi_{a} - \omega_{a}$, $ \bigtriangleup_{b}=  \varpi_{b}-\omega_{b} $ are the detunings between the corresponding cavity resonances frequencies $ \varpi_{a}, \varpi_{b} $ and intracavity field frequencies $\omega_{a}$, $\omega_{b}$, respectively. L  is the cavity length, $ \kappa=\dfrac{\sqrt{2}d_{eff}K}{cn_{a}\sqrt{\varepsilon_{0}cn_{b}S}} $is the nonlinear  coupling coefficient, where $ d_{eff} $ is the nonlinear coefficient, $ K= sinc(\bigtriangleup \beta L/2) $ is a dimensionless factor which related to the phase mismatch $ \bigtriangleup \beta $.  $ n_{a} $, $ n_{b} $ are effective refractive indices of the field a, b. c is the speed of light, $ \varepsilon_{0} $ is the vacuum permittivity, S is mode overlap area. $ \eta = \theta_{a}/t_{R} $ represents the average coupling rate
of $ \omega_{a} $ filed. $ t_{R} $ is the roundtrip time and $ \theta_{a} $ is the responding coupling rate.
\begin{figure}[tb]
	\centering
	\includegraphics[width=\linewidth]{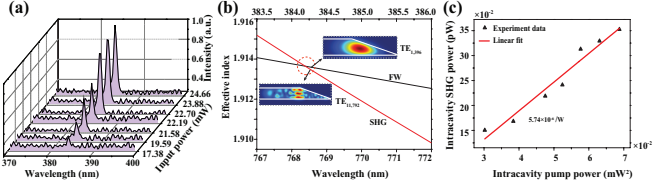}
	\caption{ 
		UV second-harmonic generation in the LTOI microdisk. (a) The recorded spectrum
		of the UV SHG signal (around 384.3 nm) with different input power.  (b) Effective indices as functions of wavelength of TE$_{1,396}$ in the visible band and TE$_{11,792}$ in the UV band, Insets show the simulated mode profiles of the FW and UV SH waves. (c) Intracavity pump power dependency of the generated SH power.}
	\label{Figure3}
\end{figure}
Considering the undepleted-pump approximation, we can obtained the power conversion efficiency from the coupled-mode equation as,
\begin{eqnarray}
	\xi=\dfrac{64\omega_{a}^{2}\kappa^{2}\eta^{2}t_{R}\theta_{b}}{(4\bigtriangleup_{a}^{2}+\alpha_{a}^{2})^{2}[\alpha_{c}^{2}+4(2\bigtriangleup_{a}+\bigtriangleup_{b}^2)]}P_{in}
\end{eqnarray}
where $ P_{in}=|a_{in}^{2}|$ is the pump power. The ratio of $ \xi/P_{in} $ is related to the phase matching condition, the propagation losses, and the mode profiles. For fixed propagation losses, 
the highest efficiency can be achieved when satisfied the perfect phase matching and critically coupled conditions.
\begin{table*}[hp!t]
	\caption[
	]{Comparison of optical parameters of lithium tantalate (LiTaO$_{3}$, LT) and lithium niobate (LiNbO$_{3}$, LN)[45,68]}
	\label{tab1}
	\centering
	\begin{tabular}{@{}lll@{}} \toprule
		\hline
		\centering
		
		Crystals   & LiNbO$_{3}$(LN) & LiTaO$_{3}$(LT)\\ \midrule
		\hline
		Point group & 3m & 3m \\
		\hline
		Transparency range \\at '0' transmittance level & 0.40-5.5 $\mu$m & 0.28-5.5 $\mu$m \\
		\hline
		Second-order nonlinear efficients & $ |d_{33}| $  = 34.1 pm/V   & $ |d_{33}| $ = 26.2 pm/V  \\
		\hline
		Linear absoption coefficient &  0.0019-0.0023 /cm   & 0.001-0.003/cm \\
		\hline
		Laser-induced surface & 0.005-0.03 GW/cm$^{2}$  & 0.22GW/cm$^{2}$ \\
		damage threshold  &(@1064 nm and 10 ns)  & (@1060 nm and 30 ns) \\
		\hline
		Photorefractive damage \\threshold @ 532 nm & 1 KW/cm$^{2}$  &  2000 KW/cm$^{2}$ \\
		\hline
		Coercive field value & $ \sim$ 21 KV/mm & $ \sim$ 21 KV/mm \\
		\hline
		Linear electrooptic \\coefficients @632.8 nm & $ |\gamma_{33} |$=30.8 pm/V & $ |\gamma_{33}| $=30.3 pm/V \\ \bottomrule
		\hline
		
	\end{tabular}
\end{table*}

We calculated the effective indices as a function of the input wavelength in visible and UV.
Fig. 7(b) shows the effective refractive indices of the input pump light in the visible band and the SHG signal in the UV band, which gives a phase-matched pump wavelength around 768.50 nm (red dotted circle) at the temperature of 25 °C. Insets in Fig. 7(b) show the simulated mode profile of the fundamental wave (FW) and UV second harmonic, respectively. According to the mode profile of FW and UV second harmonic, the mode phase matching is satisfied in the LTOI microdisk.
The power-dependent performance of UV SHG on the visible input at 768.60 nm is shown in Fig. 7(c). The quadratic correlation between the pump light and UV SH signal power, well confirms that the UV signal is generated by SHG of the LTOI microdisk originating from the $\chi^{(2)}$
nonlinearity.
Fig. 7(c) also shows the observed intracavity SHG power as a function of the intracavity pump power square, and the normalized conversion efficiency is 5.74$ \times$ 10$ ^{-6} $/W. 
Considering in our experiment, the same tapered fiber was used to couple in the input pump light and out the UV signal, and the generated UV signal was filtered by a spatial filter before being received into the spectrometer. The actually generated UV signal and conversion efficiency should be larger, and to optimize the coupling and filtering schemes, and design the domain structure carfully etc., the conversion efficiency can be improved higher.

In conclusion, we fabricated 50-$\mu$m-diameter microdisk resonators on a Z-cut LTOI which have high Q factors both in the telecom and visible band.
By making use of the large input pump power in the telecom band, the near-infrared and visible (red, green) light are also experimentally achieved, which originate from the cFWM, and cSFG/THG in the LTOI microdisk, respectively. 
To the best of our knowledge, it is the first time to realize  $ \chi^{(2)} $- $ \chi^{(2)} $ and  $ \chi^{(2)} $-$ \chi^{(3)} $-based nonlinear frequency generation simultaneously in an LTOI microresonator.
What's more, combining the relatively large nonlinear coefficients and UV transparency of LTOI, we also achieved on-chip efficient long-wave UV (UV-A, 320 to 400 nm) radiation by using the SHG process.
Such on-chip UV light achieved by the small device size is beneficial to integrating and generating a more compact UV source. Moreover, one can obtain ultraviolet light to the medium-wave UV (UV-B, 280 to 320 nm) by optimizing the pump wavelength, power, and phase-matching conditions.
More optical nonlinear processes, for example, optical parametric oscillators (OPOs) and difference frequency generation (DFG), can be generated by designing periodically poled domain structures and the device dispersion in the LTOI microdisk. More broad spectral span frequency combs may also be achieved by careful dispersion design combining the high damage threshold. In addition, considering the similar physical properties of LT and LN (shown in Table 1), we can also apply a voltage regulation to obtain tunable nonlinear frequency output. To make use of the advantages of LT, one may even direct heterobonding it with LN in further studies [45,75,79]. 
All these nonlinear process generations in the LTOI microdisk show the great application potential for LTOI in integrated nonlinear photonics and its excellent physical properties make it can be an excellent nonlinear photonics and hybrid photonics platform.

This work was supported by the National Key R $\&$ D Program of China (Grant Nos.  2019YFB2203501, and  2018YFA0306301), the National Natural Science Foundation of China (Grant Nos. 12134009, 91950107), Shanghai Municipal Science and Technology Major Project (2019SHZDZX01-ZX06), and SJTU No. 21X010200828.
	
The authors declare no conflicts of interest.

Data underlying the results presented in this paper are not publicly available at this time but may be obtained from the authors upon reasonable request.

* These authors contributed equally to this work.

$ \star $ The data that support the findings of this study are available from the corresponding author upon reasonable request.


%
%

%


1  J. Y. Chen, C. Tang, Z. H. Ma, Z. Li, Y. M. Sua, and Y. P. Huang, Optics Letters 45, 3789–3792 (2020).\\
2  R. Luo, Y. He, H. Liang, M. Li, and Q. Lin, Optica 5, 1006–1011 (2018).\\
3  Y. Zhao, J. K. Jang, Y. Okawachi, and A. L. Gaeta, Optics Letters 46, 5393–5396 (2021).\\
4  C.-Q. Xu and B. Chen, Optics Letters 29, 292–294 (2004).\\
5  L. Ledezma, R. Sekine, Q. Guo, R. Nehra, S. Jahani, and A. Marandi, Optica 9, 303–308 (2022).\\
6  C. Wang, C. Langrock, A. Marandi, M. Jankowski, M. Zhang, B. Desiatov, M. M. Fejer, and M. Loncar, Optica 5, 1438–1441 (2018).\\
7  D. Puzyrev, V. Pankratov, A. Villois, and D. Skryabin, Physical Review A 104, 013520 (2021).\\
8  J.-Q. Wang, Y.-H. Yang, M. Li, X.-X. Hu, J. B. Surya, X.-B. Xu, C.-H. Dong, G.-C. Guo, H. X. Tang, and C.-L. Zou, Physical Review Letters 126, 133601 (2021).\\
9  R. W. Boyd, Nonlinear optics (Academic press, 2020).\\
10  K. J. Vahala, Nature 424, 839–846
(2003).\\
11  J. Chen, X. Shen, S. Tang, Q. Cao, Q. Gong, and Y. Xiao, Physical Review Letters 123, 173902 (2019).\\
12  P.-J. Zhang, Q.-X. Ji, Q.-T. Cao, H. Wang, W. Liu, Q. Gong, and Y.-F. Xiao, Proceedings of the National Academy of Sciences 118, e2101605118 (2021).\\
13  L. Yao, P. Liu, H.-J. Chen, Q. Gong, Q.-F. Yang, and Y.-F. Xiao, Optica 9, 561–564 (2022).\\
14  B.-Y. Xu, L.-K. Chen, J.-T. Lin, L.-T. Feng, R. Niu, Z.-Y. Zhou, R.-H. Gao, C.-H. Dong, G.-C. Guo, Q.-H. Gong, et al.,  Science China Physics, Mechanics \& Astronomy 65, 1–7 (2022).\\
15  Y.-Y. Li, Q.-T. Cao, J.-h. Chen, X.-C. Yu, and Y.-F. Xiao, Physical Review Applied 16, 044016 (2021).\\
16  X.-C. Yu, S.-J. Tang, W. Liu, Y. Xu, Q. Gong, Y.-L. Chen, and Y.-F. Xiao, Proceedings of the National Academy of Sciences 119, e2108678119 (2022).\\
17  D. Xu, Z. Z. Han, Y. K. Lu, Q. Gong, C. Qiu, G. Chen, and Y. Xiao, Advanced Photonics 1, 046002 (2019).\\
18  J.-C. Shi, Q.-X. Ji, Q.-T. Cao, Y. Yu, W. Liu, Q. Gong, and Y.-F. Xiao, Physical Review Letters 128, 073901 (2022).\\
19  J. Lu, J. B. Surya, X. Liu, A. W. Bruch, Z. Gong, Y. Xu, and H. X. Tang, Optica 6, 1455–1460 (2019).\\
20  I. Breunig, Laser \& Photonics Reviews 4, 569–587 (2016).\\
21  R. Zhuang, J. He, Y. Qi, \& Y. Li,  Advanced Materials, 2208113 (2022).\\
22  C. Wang, M. Zhang, X. Chen, M. Bertrand, A. Shams-Ansari, S. Chandrasekhar, P. Winzer, and M. Loncar, Nature 562, 101–104 (2018).\\
23  M. He, M. Xu, Y. Ren, J. Jian, Z. Ruan, Y. Xu, S. Gao, S. Sun, X. Wen, L. Zhou, and orthers, Nature Photonics 13, 359–364 (2019).\\
24  Z. Wang, G. Chen, Z. Ruan, R. Gan, P. Huang, Z. Zheng, L. Lu, J. Li, C. Guo, K. Chen, and L. Liu, ACS Photonics (2022).\\
25  M. Xu, M. He, H. Zhang, J. Jian, Y. Pan, X. Liu, L. Chen, X. Meng, H. Chen, Z. Li, and Y. S. Y. S. C. Xinlun, Nature Communications 11, 1–7(2020).\\
26  C. Wang, M. Zhang, M. Yu, R. Zhu, H. Hu, and M. Loncar, Nature Communications 10, 1–6(2019).\\
27  M. Zhang, B. Buscaino, C. Wang, A. Shams-Ansari, C. Reimer, R. Zhu, J. M. Kahn, and M. Loncar, Nature 568, 373–377 (2019).\\
28  Z. Hao, J. Wang, S. Ma, W. Mao, F. Bo, F. Gao, G. Zhang, and J. Xu, Photonics Research 5, 623–628(2017).\\
29  J. Lin, N. Yao, Z. Hao, J. Zhang, W. Mao, M. Wang, W. Chu, R. Wu, Z. Fang, L. Qiao, et al., Physical Review Letters 122, 173903(2019).\\
30  L. Ge, Y. Chen, H. Jiang, G. Li, B. Zhu, X. Chen, et al., Photonics Research 6, 954–958 (2018).\\
31  S. Liu, Y. Zheng, and X. Chen, Optics Letters 42, 3626–3629 (2017).\\
32  R. Luo, Y. He, H. Liang, M. Li, J. Ling, and Q. Lin, Physical Review Applied 11, 034026(2019).\\
33  H. Jiang, H. Liang, R. Luo, X. Chen, Y. Chen, and Q. Lin, Applied Physics Letters 113, 021104 (2018).\\
34  Y. Li, Z. Huang, W. Qiu, J. Dong, H. Guan, and H. Lu, Chinese Optics Letters 19, 060012(2021).\\
35  F. Ye, Y. Yu, X. Xi, and X. Sun, Laser \& Photonics Reviews 16, 2100429 (2022).\\
36  C. Wang, C. Langrock, A. Marandi, M. Jankowski, M. Zhang, B. Desiatov, M. M. Fejer, and M. Loncar, Optica 5, 1438–1441 (2018).\\
37  J.-Y. Chen, Z.-H. Ma, Y. M. Sua, Z. Li, C. Tang, and Y.-P. Huang, Optica 6, 1244–1245(2019).\\
38  N. Amiune, D. N. Puzyrev, V. V. Pankratov, D. V. Skryabin, K. Buse, and I. Breunig, Optics Express 29, 41378–41387(2021).\\
39  J. Fan, J. Zhao, L. Shi, N. Xiao, and M. Hu, Advanced Photonics 2, 045001 (2020).\\
40  H. Jin, F. M. Liu, P. Xu, J. L. Xia, M. L. Zhong,  Y. Yuan, S. N. Zhu, Physical Review Letters, 113(10), 103601 (2014).\\
41  Y. Liu, X. Yan, J. Wu, B. Zhu, Y. Chen, and X. Chen, Science China Physics, Mechanics \& Astronomy 64, 1–5 (2021).\\
42  Z. Chen, Q. Xu, K. Zhang, W.-H. Wong, D.-L. Zhang, E. Y.-B. Pun, and C. Wang, Optics Letters 46, 1161–1164(2021).\\
43  X. Liu, X. Yan, H. Li, Y. Chen, X. Chen, et al., Optics Letters 46, 5505–5508 (2021).\\
44  J. Zhou, Y. Liang, Z. Liu, W. Chu, H. Zhang, D. Yin, Z. Fang, R. Wu, J. Zhang, W. Chen, et al., Laser  \&  Photonics Reviews 15, 2100030 (2021).\\
45  Q. Luo, C. Yang, Z. Hao, R. Zhang, D. Zheng, F. Bo, Y. Kong, G. Zhang, and J. Xu, Chinese Optics Letters 19, 060008 (2021).\\
46  M. Cai, K. Wu, J. Xiang, Z. Xiao, T. Li, C. Li, and J. Chen, IEEE Journal of Selected Topics in Quantum Electronics 28, 1–8 (2021).\\
47  D. N. Nikogosyan, Nonlinear optical crystals: a complete survey (Springer Science \& Business Media, 2006).\\
48  K. Moutzouris, G. Hloupis, I. Stavrakas, D. Triantis, and M.-H. Chou, Optical Materials Express 1, 458–465 (2011).\\
49  L. Wang, X. Zhang, L. Li, Q. Lu, C. Romero, J. R. V. deAldana, and F. Chen, Optics Express 27, 2101–2111 (2019).\\
50  M. Lines,  Physical Review B 2, 698 (1970).\\
51  P. J. Matthews and A. R. Mickelson, Journal of Applied Physics 71, 5310–5317 (1992).\\
52  P. J. Matthews, A. R. Mickelson, and S. W. Novak, Journal of Applied Physics 72, 2562–2574 (1992).\\
53  B. Wu, B. Zhang, W. Liu, Q. Lu, L. Wang, and F. Chen, Optics \& Laser Technology 145, 107500 (2022).\\
54  Z. Gao, S. Zhu, S.-Y. Tu, and A. Kung, Applied Physics Letters 89, 181101 (2006).\\
55  A. Busacca, E. D Asaro, A. Pasquazi, S. Stivala, and G. Assanto, Applied Physics Letters 93, 121117 (2008).\\
56  M. Katz, R. Route, D. Hum, K. Parameswaran, G. Miller, and M. Fejer, Optics Letters 29, 1775–1777 (2004).\\
57  M. Klein, D.-H. Lee, J.-P. Meyn, B. Beier, K.-J. Boller, and R. Wallenstein, Optics Letters 23, 831–833 (1998).\\
58  S. C. Kumar, M. Ebrahim-Zadeh, et al., Optics Letters 44, 5796–5799 (2019).\\
59  Y. Liu, Z. Xie, W. Ling, Y. Yuan, X. Lv, J. Lu, X. Hu, G. Zhao, and S. Zhu, Optics Express 19, 17500–17505 (2011).\\
60  J.-P. Meyn and M. Fejer, Optics Letters 22, 1214–1216 (1997).\\
61  V. Y. Shur, A. R. Akhmatkhanov, M. A. Chuvakova, A. A. Esin, O. L. Antipov, A. A. Boyko, and D. B. Kolker, in The European Conference on Lasers and Electro-Optics (Optical Society of America, 2019) p. ce p 36.\\
62  J.-P. Meyn, C. Laue, R. Knappe, R. Wallenstein, and M. Fejer, Applied Physics B 73, 111–114 (2001).\\
63  S. V. Tovstonog, S. Kurimura, and K. Kitamura, Applied Physics Letters 90, 051115 (2007).\\
64  H. Xie, W.-Y. Hsu, and R. Raj, Journal of Applied Physics 77, 3420–3425 (1995).\\
65  Z. Yellas, M. W. Lee, R. Kremer, K.-H. Chang, M. R. Beghoul, L.-H. Peng, and A. Boudrioua, Optics Express 25, 30253–30258 (2017).\\
66  N. E. Yu, S. Kurimura, Y. Nomura, M. Nakamura, K. Kitamura, J. Sakuma, Y. Otani, and A. Shiratori, Applied Physics Letters 84, 1662–1664 (2004).\\
67  P. Xu, S. Ji, S. Zhu, X. Yu, J. Sun, H. Wang, J. He, Y. Zhu, and N. Ming,  Physical Review Letters 93, 133904 (2004).\\
68  S. N. Zhu, Y. Y. Zhu, and N. B. Ming, Science 278, 843–846 (1997).\\
69  H. Leng, X. Yu, Y. Gong, P. Xu, Z. Xie, H. Jin, C. Zhang, and S. Zhu, Nature Communications 2, 1–5 (2011).\\
70  P. Dittrich, B. Koziarska-Glinka, G. Montemezzani, P. Gunter, S. Takekawa, K. Kitamura, and Y. Furukawa, Journal of the Optical Society of America B 21, 632–639 (2004).\\
71  J. Imbrock, S. Wevering, K. Buse, and E. Kratzig, JOSA B 16, 1392–1397 (1999).\\
72  M. Soltani, V. Ilchenko, A. Matsko, A. Savchenkov, J. Schlafer, C. Ryan, and L. Maleki, Optics Letters 41, 4375–4378 (2016).\\
73  M. Jacob, J. Hartnett, J. Mazierska, J. Krupka, and M. Tobar, in TENCON 2003. Conference on Convergent Technologies for Asia-Pacifc Region, Vol. 4(IEEE, 2003) pp. 1362–1366.\\
74  J. Koskela, J. V. Knuuttila, T. Makkonen, V. P. Plessky, and M. M. Salomaa, IEEE Transactions on Ultrasonics, Ferroelectrics, and Frequency Control 48, 1517–1526 (2001).\\
75  R. Denton, F. Chen, and A. Ballman, Journal of Applied Physics 38, 1611–1617 (1967).\\
76  X. Lu, G. Moille, A. Rao, D. A. Westly, and K. Srinivasan, Optica 7, 1417–1425 (2020).\\
77  M. Li,  C. L. Zou, C. H. Dong ,X. F. Ren,2 and D. X. Dai, Physical Review A, 2018, 98(1): 013854.\\
78  M. Li, C.-L. Zou, C.-H. Dong, and D.-X. Dai, Optics Express 26, 27294–27304 (2018).\\
79  K. Eda, M. Sugimoto, and Y. Tomita, Applied Physics Letters 66, 827–829 (1995).
\end{document}